\documentstyle[prl,aps,epsf,epsfig,twocolumn]{revtex}
\begin{document}
\draft
\title{Classical invariants and the quantum-classical link }
\author{Diego A. Wisniacki$^{1,2}$ and Eduardo Vergini$^1$}

\address{$^1$Dep. de F\'\i sica,
   Comisi\'on Nacional de Energ\'\i a At\'omica,
   Avenida del Libertador 8250,
   1429 Buenos Aires, Argentina.}
\address{$^2$Departamento de F\'\i sica ``J.J. Giambiagi'',
   FCEN, UBA, Pabell\'on 1, Ciudad Universitaria,
   1428 Buenos Aires, Argentina.}
\date{Received \today}
\maketitle
\begin{abstract}
The classical invariants of a Hamiltonian system are expected to be derivable
from the respective quantum spectrum. In fact, semiclassical expressions relate
periodic orbits with eigenfunctions and  eigenenergies of classical chaotic
systems. Based on trace formulae, we construct smooth functions 
highly localized in the 
neighborhood of periodic orbits using only quantum information.
Those functions show how classical hyperbolic structures emerge from 
quantum mechanics
in chaotic systems. Finally, we discuss the proper quantum-classical link.

\end{abstract}
\pacs{PACS numbers: 05.45.+b, 03.65.Ge, 03.65.Sq}

For more than 20 years a lot of effort has been done in order to 
clarify the interplay between quantum and classical behavior in
chaotic systems. Periodic orbit theory shows the important role 
that classical invariants play in quantum mechanics. 
For example, the Gutzwiller trace formula relates the quantum spectral 
density with classical periodic orbits \cite{gutz}. 

Less is known about the interplay
between the neighborhood of periodic orbits (the stable and unstable
manifolds) and quantum mechanics. 
The first step was given by Bogomolny \cite{bogo} who found the 
semiclassical contribution of periodic 
orbits and their vicinities to wave functions smoothed in energy and position.
Advances in this issue 
contribute, among others, to the study of the morphologies of wave 
functions of chaotic systems. Especially, this gives new insights to the 
debated problem, the scarring phenomena 
\cite{Heller84}, an anomalous 
localization of quantum probability density along unstable periodic 
orbits in classically chaotic systems.
For example,  Kaplan and Heller \cite{KaplanX} proposed test 
states constructed  by  coherent wave-packet sums centered on a periodic orbit. 
The constructed wave function lives not only on the periodic orbit, but also along 
the linearized invariant manifolds. On the other hand, Vergini and Carlo
\cite{vergi1,vergi2}  
developed a semiclassical construction of wave functions on periodic
orbits and their neighborhoods (the stable and unstable manifolds)
which are highly localized in energy.
To resemble the classical hyperbolic structure, these constructions 
use the orbit and 
the linearized dynamics around it. The question that naturally  arises is if 
the classical hyperbolic structure is present in quantum mechanics. 
An important contribution in this direction was given by 
Nonnenmacher and Voros \cite{voros}. They studied the eigenstates 
structures around a 
hyperbolic point in a one dimensional system.

In this letter,
we address that question by constructing smoothed functions living on the
periodic orbit and their manifolds and we use only the quantum information 
(the eigenvalues and eigenfunctions) of the system . 
This clearly will show the skeleton on which the quantum mechanics is
built. Moreover, some features of these smoothed functions 
confirm predictions by Bogomolny and are 
well understood with the semiclassical construction 
of Ref. \cite{vergi1}.

Our starting point is the Gutzwiller trace formula \cite{gutz}.
This formula evaluates the quantal spectrum of energy of a bounded 
system in terms of the classical spectrum of periodic orbits.  
We are going to consider billiard systems where
the classical motion is described by straight lines between consecutive
bounces with the boundary. The incoming and outgoing trajectories
at a bounce satisfy specular reflection.
A periodic orbit $\delta$ is characterized by the length $L_{\delta}$, 
and the Maslov index $\nu_{\delta}$.
 The quantum mechanics is given 
by the set of eigen-wavenumbers
${k_{n}}$ and eigenfunctions ${\phi_{n} (\vec{r})}$ which are defined by the
Helmholtz equation $\;(\nabla^{2}+k_{n}^{2})\phi_{n} (\vec{r})=0\;$, with 
Dirichlet boundary conditions.  For these systems, the trace formula reads
\cite{coh}
\begin{equation}
\sum_{n} \delta(k-k_{n})=\rho(k)+\frac{1}{2 \pi} \sum_{\delta} A_{\delta} 
\;e^{-i k L_{\delta}},
\label{trace}
\end{equation}
with
\begin{equation}
A_{\delta}=\frac{L_{\delta} \;
e^{i  \nu_{\delta} \pi/2}}  
{ \sqrt{ |\det(M_{\delta}-I)|}}.
\end{equation}
and $M_{\delta}$ is the monodromy 
matrix of the orbit $\delta$.  $\rho(k)$ is the mean level density

\begin{figure}[htbp]
\centering \leavevmode
\epsfxsize=4cm
\center{\epsfig{file=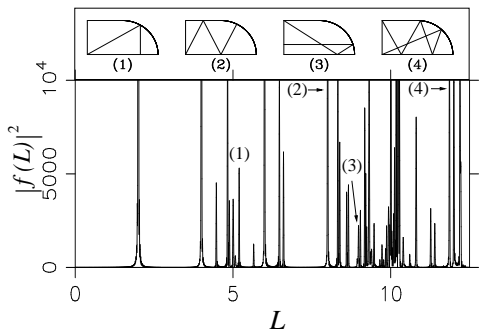, ,width=6.5cm,angle=0}}
\vspace{0.5cm}
\caption{$|f(L)|^{2}$ evaluated from the first $9910$ eigenvalues of
the desymmetrized stadium billiard with radius
$r=1$ and straight line of length $a=1$. The labels over some of the
peaks indicate the corresponding  periodic orbits 
shown in the inset.}
\label{fig1} 
\end{figure}

The Fourier transform of Eq. (\ref{trace}) gives the classical
spectrum in terms of the quantal one \cite{wintgen}
\begin{equation}
f(L) \equiv \sum_{k_{n}<k} e^{i k_{n} L}-
\tilde{\rho}(L)=
\sum_{\delta} \frac{A_{\delta}\; \sin [k(L-L_{\delta})]}{2 \pi\;
(L-L_{\delta})}.
\label{fourier}
\end{equation} 
In particular, $\;f(L_{\delta})\simeq k\;A_{\delta}/2 \pi\;$ and
$|f(L)|^{2}$ has peaks at the length of the periodic orbits.
Fig. ~\ref{fig1} shows $|f(L)|^{2}$ obtained from the eigen-wavenumbers
up to $k=264$ for the desymmetrized stadium billiard, a classical chaotic 
system \cite{buni}. The position of the peaks
agrees with the length of the orbits within an error 
($1/k$) given by the used window. The height of the peaks
and the phases are also well reproduced. 

\begin{figure}[htbp]
\centering \leavevmode
\center{\epsfig{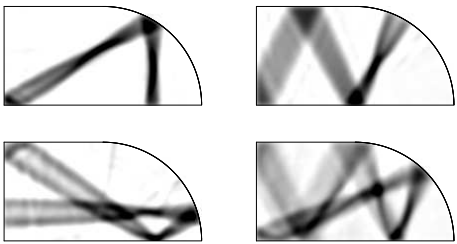}}
\vspace{0.5cm}
\caption{Spectral function (Eq. (3)) corresponding to the orbits
(1)-(4) (from upper left to bottom right) shown in the inset of 
Fig. 1. Eq. 4 was evaluated for the first
9910 eigenfunctions of the desymmetrized stadium billiard.}
\label{fig2}
\end{figure}

Phase coherent contributions occurs in Eq. (\ref{fourier}) 
when $L=L_{\delta}$. The major contribution
to the peaks is provided by the eigen-wavenumbers close to the 
Bohr--Sommerfeld quantization rule, 
$k \approx \frac{2 \pi}{L_{\nu}} 
\left( n + \frac{\nu}{4} \right)$
where $n$ is an integer. That is, the eigen-wavenumbers distribution
is not random and has information of the periodic orbits of the system
\cite{wintgen}. 
The same behavior is expected for  wave functions. Then, we propose
the following spectral function for the orbit $\delta$
\begin{equation}
\zeta_{\delta}(\vec{r}) \equiv \sum_{k_{n}<k} |\phi_{n}(\vec{r})|^{2}  
\cos( k_{n} L_{\delta}-i \pi \nu_{\delta}/2).
\label{funcion}
\end{equation}
We expect this function will retain the relevant features of the orbit 
contained in the eigenfunctions. 
In order to verify this point, we computed  the spectral functions 
(Eq. (\ref{funcion}))
for the orbits shown in the inset of Fig.~\ref{fig1} using 
the first 9910 eigenfunctions of the desymmetrized stadium 
billiard (corresponding to $0<k<264$). These spectral functions are 
shown in Fig.~\ref{fig2}. A Gaussian
smoothing with standard deviation $\sigma=0.02$ was applied in order 
to avoid the background 
random fluctuations.
Moreover, with this amount of eigenfunctions, the spectral functions of 
all the
periodic orbits up to $L \sim 14$ are well reproduced.

\begin{figure}[htbp]
\centering \leavevmode
\center{\epsfig{file=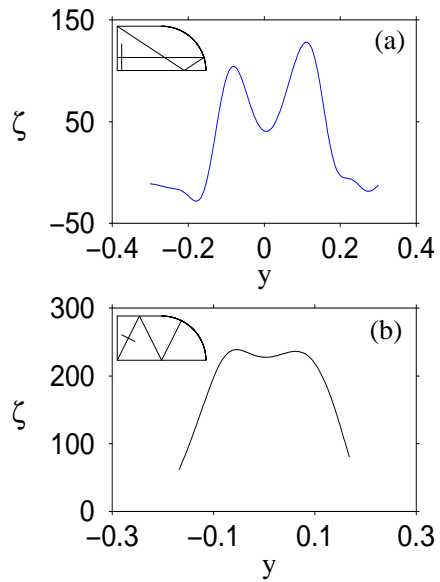, ,width=6cm,angle=0}}
\vspace{0.5cm}
\caption{Spectral function of the orbit (3) and (2) of Fig. 1 evaluated 
over the transversal section to the orbits showed in the insets.
The transversal coordinate is $y$, with $y=0$ on the orbit.}
\label{fig3}
\end{figure}

These smoothed functions are clearly localized in the neighborhood of the 
orbits and some important features are observed.
We notice that the spectral functions localize transversely to the orbits
in two different ways.  For orbits (1) and (3)
of Fig. \ref{fig1}, the maximum of the transversal localization occurs
at some distance from the orbit. Two peaks are observed
at each side of the orbit (see Fig. ~\ref{fig3} (a)). 
On the other hand, for  orbits (2) and (4) of Fig.~\ref{fig1},
the two peaks are less clear and the function $\zeta$ is high on the
orbit itself (see Fig. ~\ref{fig3} (b)). According to Bogomolny \cite{bogo}
these features depend on the characteristics of the orbit but they were not
specified. They come out clearly by comparison with the semiclassical 
construction given in Ref. \cite{vergi1}.  
From that viewpoint, these functions are a combination of the square of the 
scar functions with even and odd transversal excitations. When the classical
transversal motion is hyperbolic with reflection, the odd scar function 
quantize at the anti-Bohr energies. Then, the contribution of the odd scar 
functions to Eq. (\ref{funcion}) change their sign, producing other mixing 
with the even 
scar functions (e.g. orbits (1) an (3)). 
 
Another important characteristic observed in the spectral functions
(also predicted in Ref. \cite{bogo})
is their behavior near self-focal points of the orbit.
There, the function is enhanced and the transversal width
becomes smaller. An extra increase of the function
was also observed  at  self-crossing points of the orbit 
and in the vicinity of reflections with the boundary
 (see Fig. ~\ref{fig2}).
In a self-focal point the linearized stable 
or unstable manifold lies along the direction of the  transversal 
momentum.
This shows that the spectral functions have the imprint
of the unstable classical motion \cite{voros}.

\begin{figure}[htbp]
\centering \leavevmode
\epsfxsize=4cm
\center{\epsfig{file=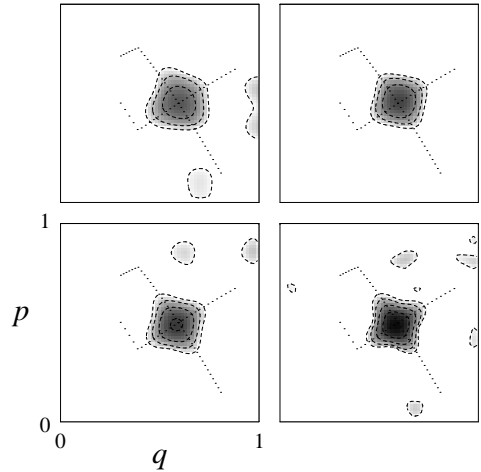, ,width=6.5cm,angle=0}}
\vspace{0.5cm}
\caption{Phase space representation of the spectral function for
orbit (1) using the first  150, 500, 2000 and 6500 eigenfunctions 
of the desymmetrized stadium billiard. The stable and unstable manifolds
are plotted with points. The coordinate $q$ is the arc length at the boundary
normalized by the perimeter of the billiard and $p$ is the fraction of tangential
momentum
$p$ .}
\label{fig4} 
\end{figure}

The last point should be clarified, therefore we have calculated 
a phase space
representation of the spectral function using the Husimi
distribution \cite{voros2} of the wave function instead
of $|\phi(\vec{r})|^2$ in Eq.(4).  As usual, we  have used 
the Birkhoff coordinates.
In the bottom right panel
of Fig. ~\ref{fig4}  the phase space representation
of the spectral function of the orbit (1) is showed.  
The stable and unstable manifolds of the
orbit are also plotted. It is observed clearly that the spectral
function lives along the manifolds.
Furthermore, it is interesting to determine how this hyperbolic
structure emerges in the semiclassical limit. This limit
occurs when  $\hbar \rightarrow 0$
or $k \rightarrow \infty$. Figure ~\ref{fig4} shows 
the phase space representation of the spectral function of orbit
(1) using the first 150, 500, 
2000 and 6500 eigenfunctions of the system. 
Increasing the number of eigenfuntions, as we are going to 
the semiclassical
limit, the  functions
become more localized over the manifolds.

Finally, there is an important point to be considered.
A remarkable aspect of Eq. (\ref{fourier}) [or (\ref{funcion})] 
is that it only permits to obtain periodic orbits with length 
satisfing
$L \leq  \ln(k L)/h$ (with $h$ the 
topological entropy).
This is because the width of the defined peaks is $\sim 1/k$
and the density of
peaks increases exponentially with $L$ ($\sim \exp (h L)/L$).
On the other hand, in order to recover the quantum spectrum with the trace 
formula [Eq.(\ref{trace})], we need all the periodic orbits with length up to 
the Heisenberg length $L_{H}=k A $ (with $A$ the area of the billiard).
In this way, we arrive to an asymmetrical situation where the classical information
obtained from the quantum spectrum is wholly insufficient to recover the initial 
quantum information.
This apparent contradiction was recently showed up with the development of a semiclassical theory 
of short periodic orbits \cite{vergi2}. In it, the length of the required periodic orbits 
(to obtain the eigenvalues and eigenfunctions up to $k$)
is lower than 
$L \leq  \ln (L_{H} h)/h=\ln (k A h)/h$.
So, we can go from quantum mechanics to classical one and vice versa 
without loss of information, and the quantum-classical link is properly established.

In conclusion, we have constructed highly localized functions in the 
vicinity of periodic orbits using only quantum 
information. With this evidence, we have shown that the classical 
hyperbolic structure of unstable periodic orbits is contained in the 
eigenfunctions of the system.
\\

We are grateful to G. Carlo, M. Saraceno and 
F. Simonotti for fruitful discussions. D. A. W. acknowledges the support from CONICET.
This work was partially supported by UBACYT (TW35), APCT PICT97 03-00050-01015,
and SECYT-ECOS.


%
%

\end{document}